\documentclass[letterpaper, 10pt, conference]{ieeeconf} 
\IEEEoverridecommandlockouts 
\usepackage{graphicx}

\usepackage{indentfirst}
\usepackage{array,multirow}
\usepackage{fancyhdr}
\usepackage{float}
\usepackage{booktabs}
\usepackage{multirow}
\usepackage{here}
\usepackage{comment}
\usepackage{siunitx}
\usepackage{hyperref}
\usepackage{algorithmicx,algorithm,algpseudocode}
\usepackage{amsmath,amssymb,amsfonts,bm,gensymb}
\usepackage{fontawesome5}
\usepackage{caption}
\usepackage{setspace}
\usepackage{xcolor}
\usepackage{cite}

\newcommand{\etal}{et~al.\ }
\newcommand{\eg}{e.\,g.,\ }
\newcommand{\ie}{i.\,e.,\ }

\title{\LARGE \bf
Motion Sickness Modeling with Visual Vertical Estimation\\and Its Application to Autonomous Personal Mobility Vehicles\vspace{-4mm}
}

\author{
Hailong~Liu$^{1}$, Shota~Inoue$^{1}$ and Takahiro~Wada$^{1}$
\thanks{$^{1}$Hailong Liu, Shota~Inoue and Takahiro Wada with the Graduate School of Science and Technology, Nara Institute of Science and Technology, 8916-5 Takayama-cho, Ikoma, Nara 630-0192, JAPAN. \faIcon[regular]{envelope}~:~{\tt\small \{liu.hailong; inoue.shota.iu1; t.wada\}@is.naist.jp}}%
}

\begin{document}
\thispagestyle{empty}
\pagestyle{empty}
\maketitle

\begin{abstract}
Passengers (drivers) of level 3--5 autonomous personal mobility vehicles~(APMV) and cars can perform non-driving tasks, such as reading books and smartphones, while driving.
It has been pointed out that such activities may increase motion sickness.
Many studies have been conducted to build countermeasures, of which various computational motion sickness models have been developed. 
Many of these are based on subjective vertical conflict~(SVC) theory, which describes vertical changes in direction sensed by human sensory organs vs. those expected by the central nervous system. 
Such models are expected to be applied to autonomous driving scenarios. 
However, no current computational model can integrate visual vertical information with vestibular sensations. 

We proposed a 6~DoF~SVC--VV model which add a visually perceived vertical block into a conventional six-degrees-of-freedom SVC model to predict VV directions from image data simulating the visual input of a human.
Hence, a simple image-based VV estimation method is proposed.

As the validation of the proposed model, this paper focuses on describing the fact that the motion sickness increases as a passenger reads a book while using an AMPV, assuming that visual vertical~(VV) plays an important role. 
In the static experiment, it is demonstrated that the estimated VV by the proposed method accurately described the gravitational acceleration direction with a low mean absolute deviation.
In addition, the results of the driving experiment using an APMV demonstrated that the proposed 6~DoF~SVC--VV model could describe that the increased motion sickness experienced when the VV and gravitational acceleration directions were different.
\end{abstract}

\section{INTRODUCTION}
\subsection{Background}
Level 3--5 autonomous driving systems~\cite{SAE_j3016b_2018} are being applied not only to cars, but also to miniaturized personal mobility vehicles~(PMVs)~\cite{liu2020_what_timeing}.
Autonomous PMVs~(APMVs) are expected to soon be widely used in mixed traffic and shared-space scenarios~\cite{morales2017social,liu2020gaze}.

In a mixed traffic environment, there will be frequent interactions among AVs (including APMVs), pedestrians, bicycles, manually driven vehicles, and other traffic participants~\cite{li2021autonomous}.
Thus, APMVs need to take collision avoidance behavior to avoid hitting obstacles and other traffic participants, frequently.
However, the passenger on the APMV may be prone to motion sickness due to frequent collision avoidance behavior.
Additionally, drivers (passengers) are allowed to perform non-driving tasks during autonomous driving~\cite{sivak2015motion,wada2016motion,diels2016self,lihmi}, \eg reading a book~\cite{isu2014quantitative,wada2020computational}, watching video~\cite{isu2014quantitative}, and virtual-reality gaming~\cite{li2021queasy}.
However, doing so may lead to motion sickness because passengers lose their spatial awareness, inhibiting their perceptions of current self-motion and predictions of future motions~\cite{Philipp2017,meschtscherjakov2019bubble}.
In particular, motion sickness may occur with high probability when visual and vestibular system are stimulated with in-congruent information~\cite{Philipp2017}.

Based on aforementioned issues, preventing motion sickness can be considered as an important challenge to the popularity and widespread use of APMVs.
To counteract these issues, various computational models of motion sickness have been used to quantify or estimate the severity of motion sickness.

\subsection{Related works}

\begin{figure*}[htb] 
\centering 
\includegraphics[width=1\linewidth]{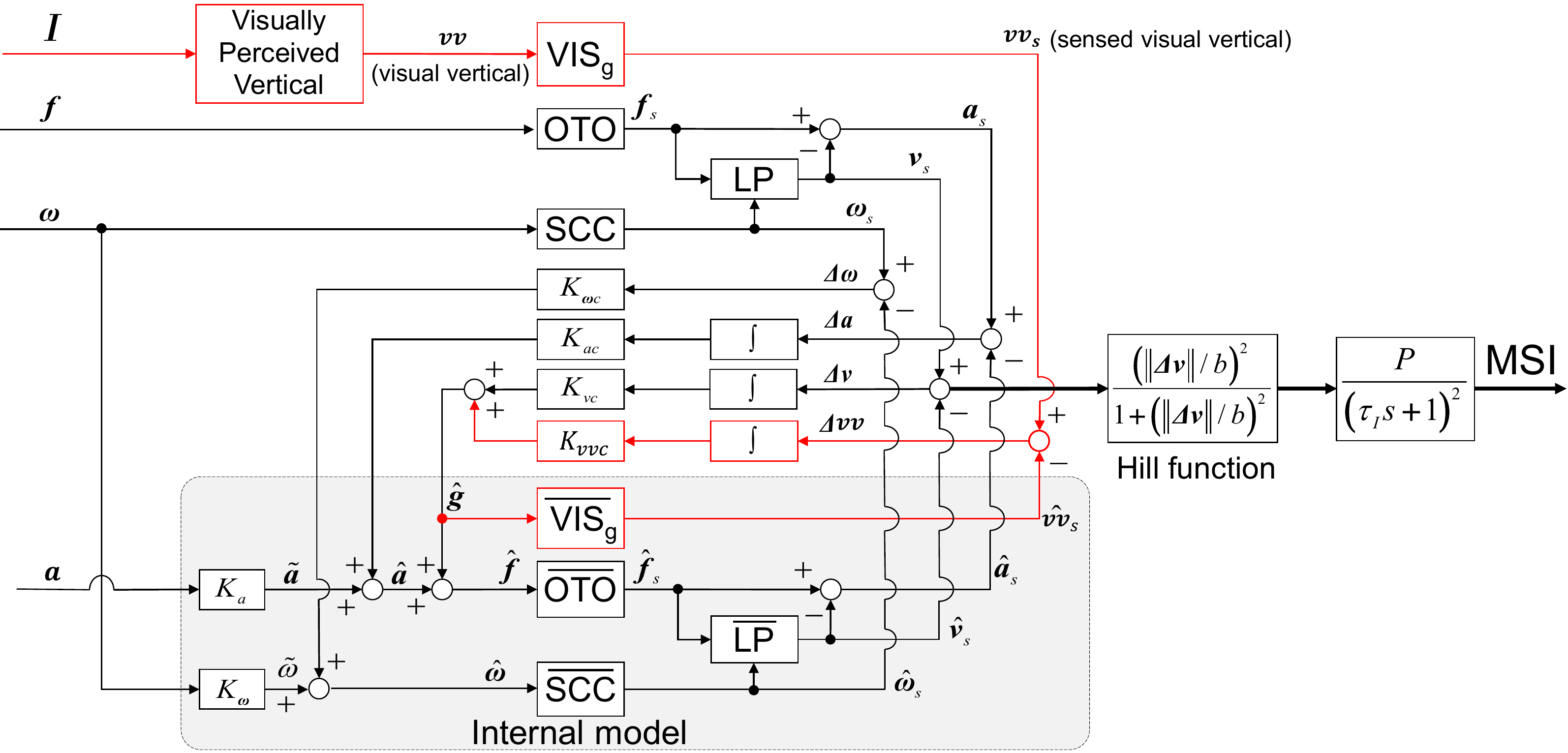}

\caption{Proposed 6~DoF SVC model including conflict of two-dimensional VV (6~DoF~SVC--VV model).} 
\label{fig:model}
\vspace{-2.5mm}
\end{figure*} 

Sensory conflict (SC) theory is widely used to explain the mechanism of motion sickness, which it postulates that motion sickness is caused by conflicts among multiple sensory signals, or between a sensory signal and anticipatory information based on previous experience~\cite{reason1978motion2}. 
Oman proposed a mathematical framework for SC theory based on observer theory, in which an internal model of sensory organs and their signals were used to generate anticipatory information~\cite{oman1990motion}.

Besides, Bles \etal proposed the subjective vertical conflict (SVC) theory~\cite{bles1998motion}, which postulated that motion sickness was mainly caused by the conflict between the vertical directions sensed by sensory organs and those estimated by their internal models.
Based on~\cite{bles1998motion}, Bos and Bles proposed the first computational model of SVC theory, detailed for one-degree-of-freedom (1~DoF) vertical motion, which used acceleration input to simulate the process of motion sickness caused by conflict between otolith organs~(OTO) and internal models~\cite{bos1998modelling}.
Kamiji \etal extended the 1~DoF SVC model~\cite{bos1998modelling} to a six-degrees-of-freedom~(6~DoF) SVC model~\cite{kamiji2007modeling}.
This model not only accepted three-dimensional (3D) acceleration input, but it also added a module of semicircular canals~(SCC) to accept 3D angular velocity input.

Efforts to include visual inputs to a SVC model have been made, such as the one by Bos \etal \cite{bos2008theory}, who proposed a motion sickness framework that included visual information such as visual angular velocities and visual vertical~(VV) information.
However, this study did not deal with a concrete method to how to apply experimental data.
As the first computational model of motion sickness that can deal with measured visual information, Wada~\etal~\cite{wada2020computational} has proposed an extended 6~DoF SVC model that calculated the angular velocities using the optical flow method from camera images to simulate human visual perception.
However, no computational model of motion sickness can deal with VV that is calculated from the measured video image.

\newpage
\subsection{Purposes and contributions}
The purposes of this paper are twofold: to propose a computational model of SVC theory that can deal with the VV estimated from video image by expanding the conventional 6~DoF SVC method~\cite{kamiji2007modeling}; 
and to demonstrate that the proposed 6~DoF~SVC--VV model describes the fact observed in a daily life scenario in which motion sickness is assumed to increase by reading a book during APMV use.

For this purpose, the contributions of this paper are:
\begin{itemize}
\item[1)] a simple image processing model is proposed to estimate the direction of two-dimensional (2D) VV.
\item[2)] a 6~DoF~SVC--VV model is proposed based on the conventional 6~DOF SVC model~\cite{kamiji2007modeling}, which represents motion sickness due to the interactions of VV from the visual system and vertical signal from the vestibular system.
\item[3)] we demonstrate that the proposed computational model has an ability to describe the fact that motion sickness is increased when reading a book while using an APMV.
\end{itemize}

\vspace{2mm}
\section{SVC MODEL WITH VV ESTIMATION}

\subsection{Conventional 6~DoF SVC model}
The conventional 6~DoF SVC model~\cite{kamiji2007modeling} is shown in part of Fig.~\ref{fig:model} as the black paths and blocks.

The input of the \framebox{\textbf{OTO}} block is the gravito-inertial acceleration (GIA), which is $\bm{f=a+g}$.
Here, $\bm{a}$ is the inertial acceleration and $\bm{g}$ is the gravitational acceleration (upward).
An identity matrix is used for transforming $\bm{f}$ to the sensed GIA $\bm{f}_s$, in \framebox{\textbf{OTO}} block. 
Its internal model \framebox{$\overline{\textbf{OTO}}$} also uses the same function with input $\bm{\hat{f}}$ which calculated from the efference copy and feedback of multiple conflicts in sensory signals.
Further, the \framebox{\textbf{SCC}} block receives the angular velocity $\bm{\omega}$ and transforms it to the sensed angular velocity $\bm{\omega}_s$ by $\bm{\omega}_s=\bm{\omega}~\tau_ds/(\tau_ds+1)$, which is the same as the function of its internal model \framebox{$\overline{\textbf{SCC}}$}.
After the \framebox{\textbf{OTO}} block and \framebox{\textbf{SCC}} block, a block \framebox{\textbf{LP}} is used to estimate the sensed vertical signal $\bm{v}_s$ by an updated law $d\bm{v}_s/dt=(\bm{f}_s-\bm{v}_s)/\tau-\bm{\omega}_s \times \bm{v}_s$.
Thereafter, the sensed inertial acceleration $\bm{a}_s$ is calculated by $\bm{a}_s=\bm{f}_s-\bm{v}_s$.

The internal model calculates the $\hat{\bm{a}}_s$, $\hat{\bm{v}}_s$, and $\hat{\bm{\omega}}_s$ with the aforementioned calculation process.
It is assumed that conflicts between the sensed signals and those expected by the internal model $\bm{a}_s$ and $\hat{\bm{a}}_s$, $\bm{v}_s$ and $\hat{\bm{v}}_s$, as well as $\bm{\omega}_s$ and $\hat{\bm{\omega}}_s$ are feedback to the internal model again to reduce the conflict. 
Please refer to~\cite{kamiji2007modeling} and \cite{wada2020computational} for the detail.

Finally, the motion sickness incidence~(MSI) is determined by
$P/(\tau_I~s+1)^2$ after the Hill function, which normalizes the conflict of vertical signals $\mathit{\Delta} \bm{v}=\bm{v}_s-\hat{\bm{v}}_s$.
Notably, the MSI represents the percentage of vomiting subjects.

\subsection{Visual vertical estimation method}
\begin{figure}[p] 
\centering 
\includegraphics[width=1\linewidth]{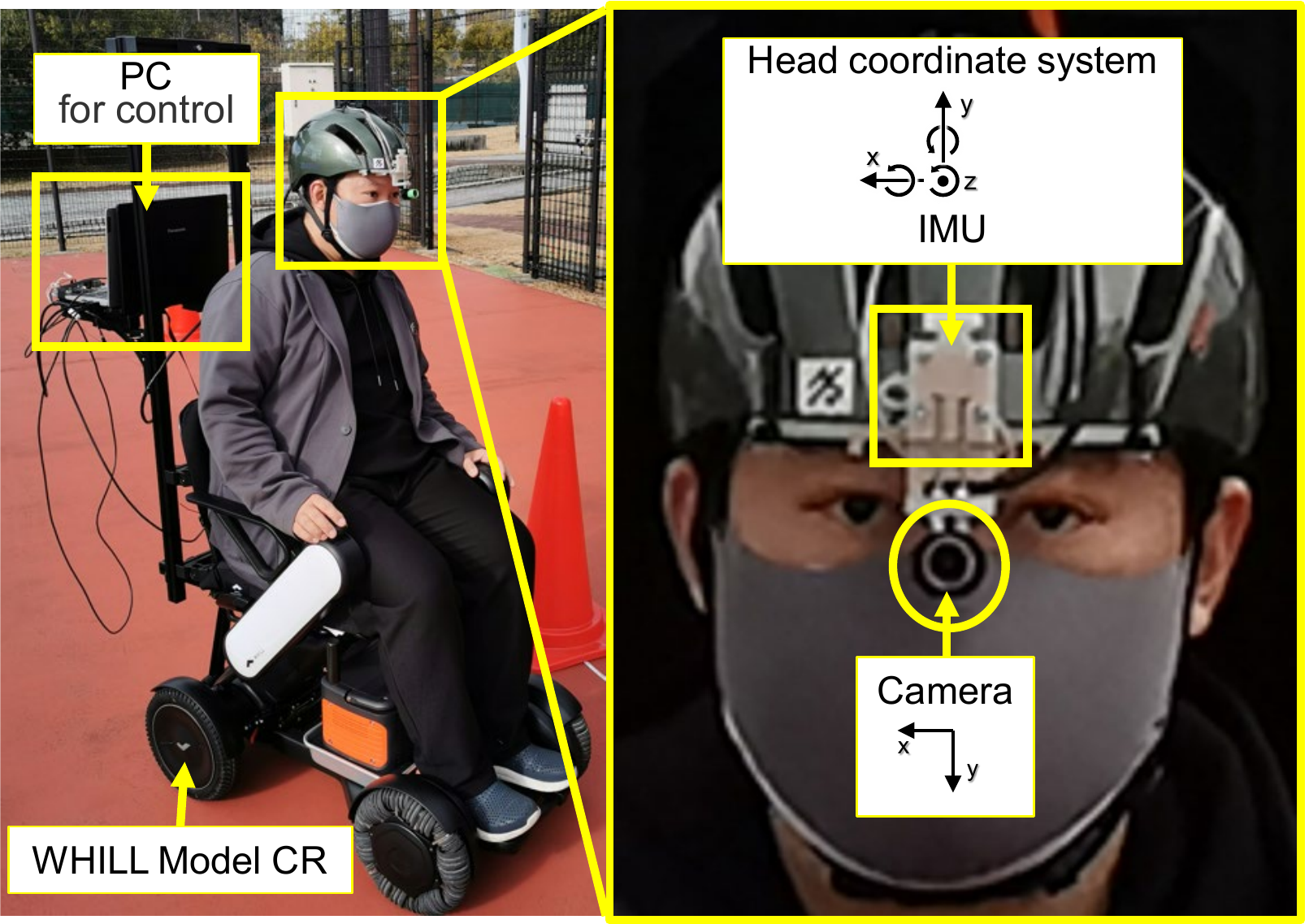}
\vspace{-6mm}
\caption{A WHILL model CR was used as the experimental vehicle. A camera and an IMU were set on a helmet to observe visual information and the acceleration as well as the angular velocity of passenger's head.} 
\label{fig:exp_whill_camera}
\vspace{3mm}
\centering 
\includegraphics[width=1\linewidth]{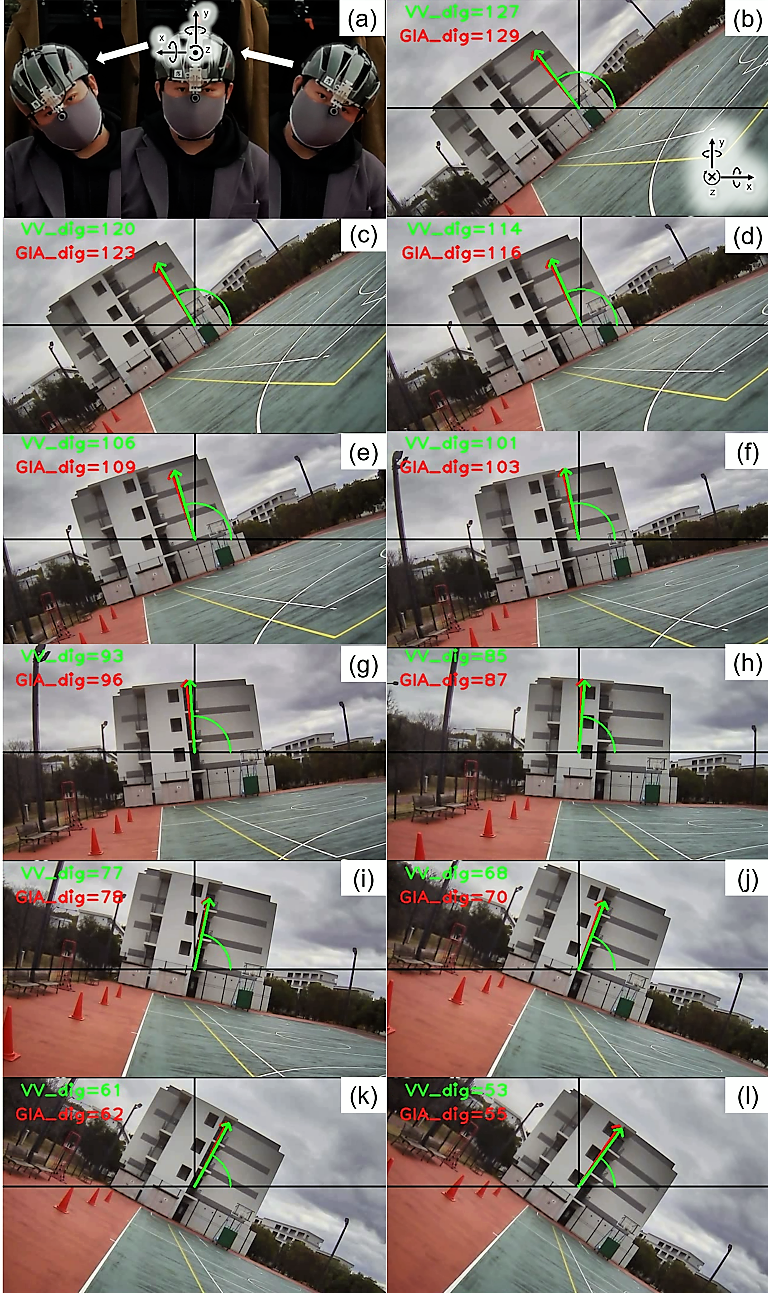}
\vspace{-6mm}
\caption{Comparison of the direction of VV and gravitational acceleration in a static state as the direction of GIA and gravitational acceleration are the same.} 
\label{fig:exp01_conditions}
\end{figure}

Since the VV is thought to be derived from cues presumed to be parallel or perpendicular to vertical objects such as buildings or the horizon in the environment~\cite{clark2019mathematical}, a simple image processing method is proposed for estimating VV by analyzing the directions of objects' edges in images.
The camera coordinate system is shown in the right part of Fig.~\ref{fig:exp_whill_camera}.

The proposed VV estimation method for \framebox{Visually Perceived Vertical} block (see Fig.~\ref{fig:model}) is shown in Algorithm~\ref{VV}.
The input is a $\bm{I}_t^{color}\in\mathbb{R}^{H\times W \times3}$ which is defined as a color image in its $t$-th frame captured by a camera attached to the human head to imitate human visual input.
Then, $\bm{I}_t^{color}$ is pre-processed by converting to a gray-scale image, reducing noise by a Gaussian filter with the size of $11 \times 11$ pixels, and normalizing through the global maximum and minimum (Algorithm~\ref{VV}, steps 1-3).

\newpage
Thereafter, Sobel operators are used to compute gradients in the transverse direction $\bm{\nabla x}_{t}$ and longitudinal direction $\bm{\nabla y}_{t}$ to detect the edges of objects in the image (Algorithm~\ref{VV}, steps 4-5).
Note that, the longitudinal axis in the camera coordinate system is downward, and the longitudinal axis in the head coordinate system is defined as upward in this paper (See the right part of Fig.~\ref{fig:exp_whill_camera}).
Therefore, the longitudinal gradient is taken as a negative value, \ie $\bm{\nabla y}_{t}$.
Gradients' magnitudes $M_t$ and angles $\Theta_t$ can be calculated from $\bm{\nabla x}_{t}$ and $\bm{\nabla y}_{t}$ by steps 6-7 in Algorithm~\ref{VV}.
Here, $\odot$ and $\oslash$ are Hadamard product and Hadamard division which were element-wise product and division.

In $\Theta_t$, we equate [360\degree,180\degree] to [0\degree,179\degree] because the angle of a person's neck usually does not exceed 180\degree (Algorithm~\ref{VV}, steps 8-12).
Besides, as the gradient's magnitude is larger, the edge is more likely.
The gradient's magnitudes $\bm{M}_{t}$ are normalized through the global maximum and minimum to $[0,1]$, and any gradient magnitude less than $0.25$ is discarded (Algorithm~\ref{VV}, steps 13-14).
Further, we used the erode algorithm~\footnote{Erode function by OpenCV 4.5: \url{https://docs.opencv.org/4.x/db/df6/tutorial_erosion_dilatation.html}} to remove the noisy and weak gradients' magnitudes (Algorithm~\ref{VV}, steps 15).

Next, histogram of gradients' angles is calculated by an indicator function from $\Theta_t$ with its weight matrix $\bm{M}_{t}$ (Algorithm~\ref{VV}, steps 16-18).
Here, the number of bins for the histogram is set to 180.
After calculating the histogram, the gradient’s angles in the range of [30\degree, 150\degree] are sorted in ascending order by their counts (Algorithm~\ref{VV}, steps 19-20).
We believe that the passenger's head will not rotate out of the range of [30\degree, 150\degree] in most driving situations.
After that, the best three angles~$\bm{\theta}^{best3}_{t}$ are selected according to the highest three counts~$\bm{c}^{best3}_{t}$ (Algorithm~\ref{VV}, steps 21-22).
Then, in the steps 23 of Algorithm~\ref{VV}, the direction of the VV at the $t$-th frame is calculated as 
\begin{eqnarray}
\theta^{vv}_t = 0.7~(\bm{\theta}^{best3}_{t} \cdot \bm{c}^{best3}_{t}+30)+0.3~\theta^{vv}_{t-1}.
\end{eqnarray}
$\theta^{vv}_t$ is also affected by the direction of the VV in the previous frame, \ie $\theta^{vv}_{t-1}$.

Finally, a 3D vector of the VV can be defined as
\begin{eqnarray}
\bm{vv}_t=[vv^x_t, vv^y_t,0]^T
\end{eqnarray}
in the head coordinate system which is shown in the right part of Fig.~\ref{fig:exp_whill_camera} (Algorithm~\ref{VV}, steps 24-26).
Here, the $vv^x_t$ and $vv^y_t$ are calculated as follows with a fixed norm $9.81~m/s^2$: 
\begin{eqnarray}
vv_t^x &=& 9.81~\cos(\theta^{vv}_t~\pi / 180)\\
vv_t^y &=& 9.81~\sin(\theta^{vv}_t~\pi / 180)
\end{eqnarray}

\begin{algorithm} [p]
\setstretch{1.25}
	\caption{Visual vertical estimation method for visually perceived vertical block in Fig.~\ref{fig:model}.} 
	\label{VV} 
	\small{\textbf{Open source}:\\ \url{https://github.com/lhl881210/Visual-Vertical-Estimation-for-6-DoF-SVC-Model}}\\
  {\bf Input:} 
  $\bm{I}^{color}_{t} \in \mathbb{R}^{H \times W \times 3}$ and $\theta^{vv}_0=90$, \\
  \hspace*{9mm} where $H=360, W=640, t\in\{1,\cdots\, T\}$\\	
  {\bf Output:} $\bm{vv}_t \in \mathbb{R}^{3}$
\begin{algorithmic} [1]
	
\State $\bm{I}^{gray}_{t} \in \mathbb{R}^{H \times W}$ $\gets$ Gray($\bm{I}^{color}$)
	
\State $\bm{I}^{gray}_{t} \in \mathbb{R}^{H \times W}$ $\gets$ Gaussianfilter($\bm{I}^{gray}_{t}$)
\State $\bm{I}^{gray}_{t} \in \mathbb{R}^{H \times W}$ $\gets$ Normalization$_{min}^{max}$($\bm{I}^{gray}_{t}$)
	
\State	$\bm{\nabla x}_{t}\in \mathbb{R}^{H \times W}$ $\gets$ Sobel$_x$($\bm{I}^{gray}_{t}$)
  
  \State $\bm{\nabla y}_{t}\in \mathbb{R}^{H \times W}$ $\gets$ $-$Sobel$_y$( $\bm{I}^{gray}_{t}$)
  \State $\bm{M}_{t}=(\bm{\nabla x}_{t}\odot\bm{\nabla x}_{t}+\bm{\nabla y}_{t}\odot\bm{\nabla y}_{t})^{\odot 1/2}$
   \State $\Theta_{t}=  (180 / \pi)\arctan(\bm{\nabla y}_{t}\oslash\bm{\nabla x}_{t})$
  \For {$i = 0$ to $H$} 
    \For {$j = 0$ to $W$}

    \State \hspace*{-4mm}$ (\Theta_{t})_{i,j}\gets
        \begin{cases}
       (\Theta_{t})_{i,j} & (0\leq(\Theta_{t})_{i,j}<180)\\
       (\Theta_{t})_{i,j} -180& (180\leq(\Theta_{t})_{i,j}<360)\\
        0 & ((\Theta_{t})_{i,j}=360)
       \end{cases}$
    \EndFor
 \EndFor

  \State $\bm{M}_{t}$ $\gets$ Normalization$_{min}^{max}$($\bm{M}_{t}$)
  
  \State $\bm{M}_{t}$ $\gets$ Threshold($\bm{M}_{t}$, 0.25)
  
  \State $\bm{M}_{t}$ $\gets$ Erode$(\bm{M}_{t},$ kernel$=[1,1,1][1,1,1]^T)$
  \For {$d = 0$ to 179}
    \State$(\bm{\theta}^{hist}_{t})_d \gets \sum_{i=0}^{H}\sum_{j=0}^{W}\bm{1}_d[(\Theta_{t})_{i,j}] (\bm{M}_{t})_{i,j}$,
    \Statex \hspace*{4mm} where $\bm{\theta}^{hist}_{t}\in\mathbb{N}^{180}$
  \EndFor
  
  \State $\bm{c}^{sort}_{t}\in \mathbb{N}^{121}\gets$Sort($(\bm{\theta}^{hist}_{t})_{29:149}$)
  
  \State $\bm{\theta}^{sort}_{t}\in \mathbb{N}^{121}\gets$ argSort$((\bm{\theta}^{hist}_{t})_{29:149})$ 
  
  \State $\bm{c}^{best3}_{t}\in \mathbb{R}^{3}\gets(\bm{c}^{sort}_{t})_{119:121}/\sum_{i=119}^{121}(\bm{c}^{sort}_{t})_i$ 

  \State $\bm{\theta}^{best3}_{t}\in \mathbb{N}^{3}\gets(\bm{\theta}^{sort}_{t})_{119:121}$ 
  
  \State $\theta^{vv}_t \gets 0.7~ (\bm{\theta}^{best3}_{t} \cdot \bm{c}^{best3}_{t}+30)+0.3~\theta^{vv}_{t-1}$ 
  
  \State $vv_t^x \gets 9.81~\cos(\theta^{vv}_t~\pi / 180)$ 	
	
  \State $vv_t^y \gets 9.81~\sin(\theta^{vv}_t~\pi / 180)$ 
  	
  \State $\bm{vv}_t	= [vv^x_t, vv^y_t,0]^T$  
 \end{algorithmic} 
\end{algorithm}

\begin{algorithm}[p]
\setstretch{1.25}
	\caption{Calculation of the sensed vertical $\bm{\hat{g}}$ (see Fig.~\ref{fig:model}) in a continuous-time system.} 
	\label{VIS_g} 
	\hspace*{\algorithmicindent}{\bf Input:} $\bm{vv}\in \mathbb{R}^{3}$\\
  \hspace*{\algorithmicindent}{\bf Output:} $\bm{\hat{g}}\in \mathbb{R}^{3}$ 
  
	\begin{algorithmic} [1]

	\State $\bm{vv}_{s}~\gets VIS_g~\bm{vv}$, where $ VIS_g=	\begin{bmatrix}
  1 & 0 &0 \\
  0 & 1 &0\\
  0 & 0 &1\\
\end{bmatrix}$ 
	
	\State $\bm{\hat{vv}}_{s}  \gets 
	\overline{VIS}_g~\bm{\hat{g}}$, where $\overline{VIS}_g=\begin{bmatrix}
  1 & 0 &0 \\
  0 & 1 &0\\
  0 & 0 &0\\
\end{bmatrix}$ 
  \State $\bm{\hat{vv}}_{s}  \gets 
	9.81~\bm{\hat{vv}}_{s} /||\bm{\hat{vv}}_{s} ||$ 
	
  \State $\bm{\hat{g}} = \bm{K}_{vvc} (\int_{0}^{t} (\bm{vv}_{s}(t)-\bm{\hat{vv}}_{s} (t)) dt +C_{0})$ 
   \Statex  \hspace*{5mm} $+\bm{K}_{vc} (\int_{0}^{t} (\bm{v}_s(t)-\bm{\hat{v}}_s(t)) dt+C_{0})$
  \end{algorithmic} 
\end{algorithm}

\subsection{Visual vertical into the 6~DoF SVC model}
The $\bm{vv}_t$ is estimated from an image, which is discrete variable. 
To input the $\bm{vv}_t$ into the 6~DoF SVC model which is a continuous-time system, the $\bm{vv}_t$ is converted as $\bm{vv}(t)$ by using a zero-order holder.
The $\bm{vv}$ is inputted to the 6~DoF SVC model and fed the visual vertical conflict back to the internal model, as shown in Fig.\ref{fig:model} by red paths and blocks.

As shown in step 1 of Algorithm~\ref{VIS_g},
the \framebox{{\textbf{VIS$_g$}}} block is added to transform the $\bm{vv}$ to the sensed visual vertical $\bm{vv}_{s}$.
For the sake of simplicity, the transform matrix is
\begin{eqnarray}
{VIS_g}=	\begin{bmatrix}      1 & 0 &0\\      0 & 1 & 0\\      0 & 0 &1      \end{bmatrix}.
\end{eqnarray}
Then, $\bm{vv}_{s}$ and $\bm{v}_s$, \ie vertical signals sensed by visual and the vestibular systems, are assumed in the same 3D space, \ie head coordinate system.
Notably, the value on the z-axis of $\bm{vv}_{s}$ should be $0$ because $\theta^{vv}$ is a rotation angle on the x-y plane of head coordinate system.

$\bm{\hat{vv}}_{s} $ represents the sensed visual vertical in the internal model, which is calculated by the \framebox{$\overline{\textbf{VIS$_g$}}$} block using the vertical sensed via visual--vestibular interaction, \ie $\hat{\bm{g}}$.
As shown in step 2 of Algorithm~\ref{VIS_g}, the transform matrix is
\begin{eqnarray}
\overline{VIS_g}=	\begin{bmatrix}      1 & 0 &0\\      0 & 1 & 0\\      0 & 0 &0      \end{bmatrix} ,
\end{eqnarray}
which projects the sensed vertical $\hat{\bm{g}}$ onto the x-y plane in head coordinate system filtering out the z-axis value.

Next, the norm of $\bm{\hat{vv}}_{s} $ is normalized to $9.81~m/s^2$ (Algorithm~\ref{VIS_g}, step 3).

Finally, as step 4 of Algorithm~\ref{VIS_g}, the sensed vertical $\bm{\hat{g}}$ can be calculated by
\begin{eqnarray}
\bm{\hat{g}} &=& \bm{K}_{vvc} (\int_{0}^{t} \mathit{\Delta} \bm{vv}(t) dt +C_{0})+\bm{K}_{vc} (\int_{0}^{t} \mathit{\Delta} \bm{v}(t) dt+C_{0})\nonumber\\
&=& \bm{K}_{vvc} (\int_{0}^{t} (\bm{vv}_{s}(t)-\bm{\hat{vv}}_{s} (t)) dt +C_{0})\nonumber \\ 
&&+ \bm{K}_{vc} (\int_{0}^{t} (\bm{v}_s(t)-\bm{\hat{v}}_s(t)) dt+C_{0}).
\end{eqnarray}
Then, $\bm{\hat{g}}$ can be updated by
\begin{eqnarray}
\frac{d\hat{\bm{g}}}{dt}=\bm{K}_{vvc}~(\bm{vv}_{s}-\bm{\hat{vv}}_{s} )+\bm{K}_{vc}(\bm{v}_s-\bm{\hat{v}}_s). 
\end{eqnarray}

\section{STATIC EXPERIMENT}
The purpose of this static experiment was to verify that the proposed VV estimation method can accurately estimate vertical direction.

\subsection{Experiment design}

To measure the acceleration and angular velocity of the passenger's head, as well as the passenger's view, an inertial measurement unit~(IMU) and a camera were set on the front of a helmet (see the right part of Fig.~\ref{fig:exp_whill_camera}).
More importantly, this helmet prevented passengers from being injured during the course of the experiment.
The IMU and camera were synchronized to $60~Hz$ by resampling, and the camera's resolution was set to $640\times360$ pixels.

An experimenter wore the helmet and sat on the stationary PMV as a passenger, bending the neck to 11 head poses at different angles on the roll axis, as shown in Fig.~\ref{fig:exp01_conditions}~(a).
The IMU and camera recorded the data while each head position was held for about $3~s$.

\newpage
\subsection{Evaluation method}
After measuring IMU data and camera data, $\theta^{vv}$ was estimated using the Algorithm~\ref{VV} from the camera data.

The GIA $\bm{f}$ measured by the IMU matched the gravitational acceleration, \ie $\bm{f}=\bm{g}$, because acceleration $\bm{a}=[0,0,0]$.
The direction of gravitational acceleration projected in the 2D head coordinate system was calculated by $\theta^{g}=180~\arctan(f_y/f_x)/\pi$.
Linear regression was used to analyze the relationship between the rotation angles of $\theta^{vv}$ and $\theta^{g}$.
Moreover, mean absolute deviation~(MAD) was used to analyze the errors between $\theta^{vv}$ and $\theta^{g}$.

\subsection{Result}

\begin{figure}[tb]
\centering 
\includegraphics[width=0.85\linewidth]{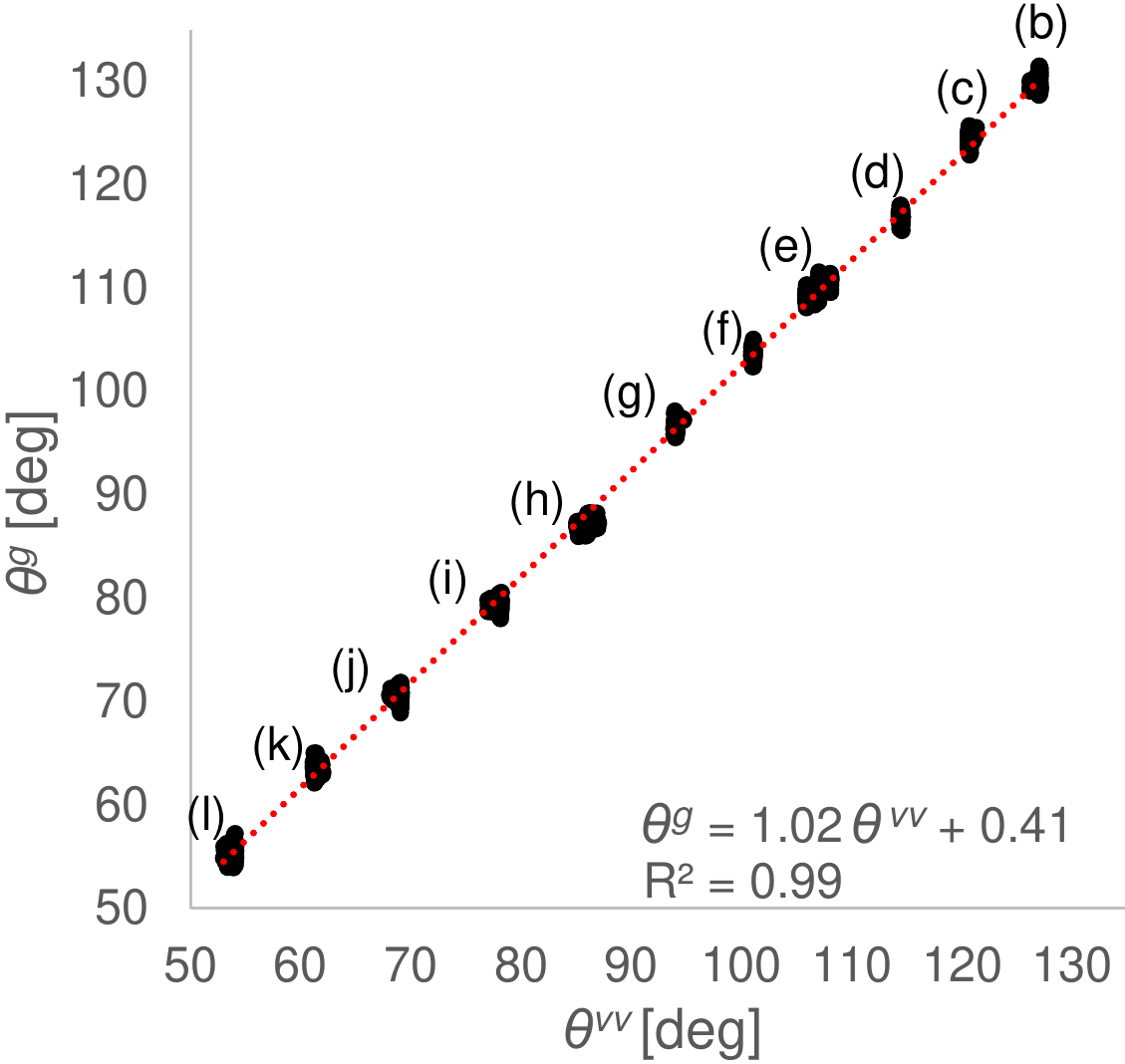}

\caption{Linear relationship between directions of VV and gravitational acceleration.}
\label{fig:exp01_result}
\vspace{-5mm}
\end{figure}

In the static experiment, a total of 1870 data frames were measured from the 11 static head poses.
Figure~\ref{fig:exp01_conditions}~(b)--(l) show images taken by the camera at different head poses, where the red and green arrows represent the gravitational acceleration (\ie GIA $\bm{f}=\bm{g}$) and VV directions, respectively.
The relationship between the directions of gravitational acceleration ($\theta^{\bm{g}}$) and VV ($\theta^{vv}$) was obtained by linear regression analysis.
The result in Fig.~\ref{fig:exp01_result} showed that $\bm{\theta}^{\bm{g}}=1.02~\bm{\theta}^{VV}+0.41$, where the coefficient of determination was $R^2=0.99$.
The MAD between $\theta^{\bm{g}}$ and $\theta^{vv}$ was $2.40$\degree.

\subsection{Discussion}
The MAD of the estimated VV in this experiment ($2.40$\degree) was within the error range of the subjective visual vertical (SVV) (\ie about $\pm 3$\degree) as reported by normal people with heads in a vertical state~\cite{guerraz2001visual,michelson2018assessment}.
Under these conditions, the proposed VV estimation method could estimate gravitational acceleration direction from image view and reached the same level as human cognition.

Notably, our method does not consider the detail of the perceived visual vertical of human, such that the range of error gradually increases with the head deflection~\cite{guerraz2001visual,michelson2018assessment}.  
Therefore, a function simulating this bias of human cognition (\ie bias of SVV) will be applied to this VV estimation method in the future.

\section{DRIVING EXPERIMENT}

\begin{figure*}[tb] 
\centering 
\includegraphics[width=0.98\linewidth]{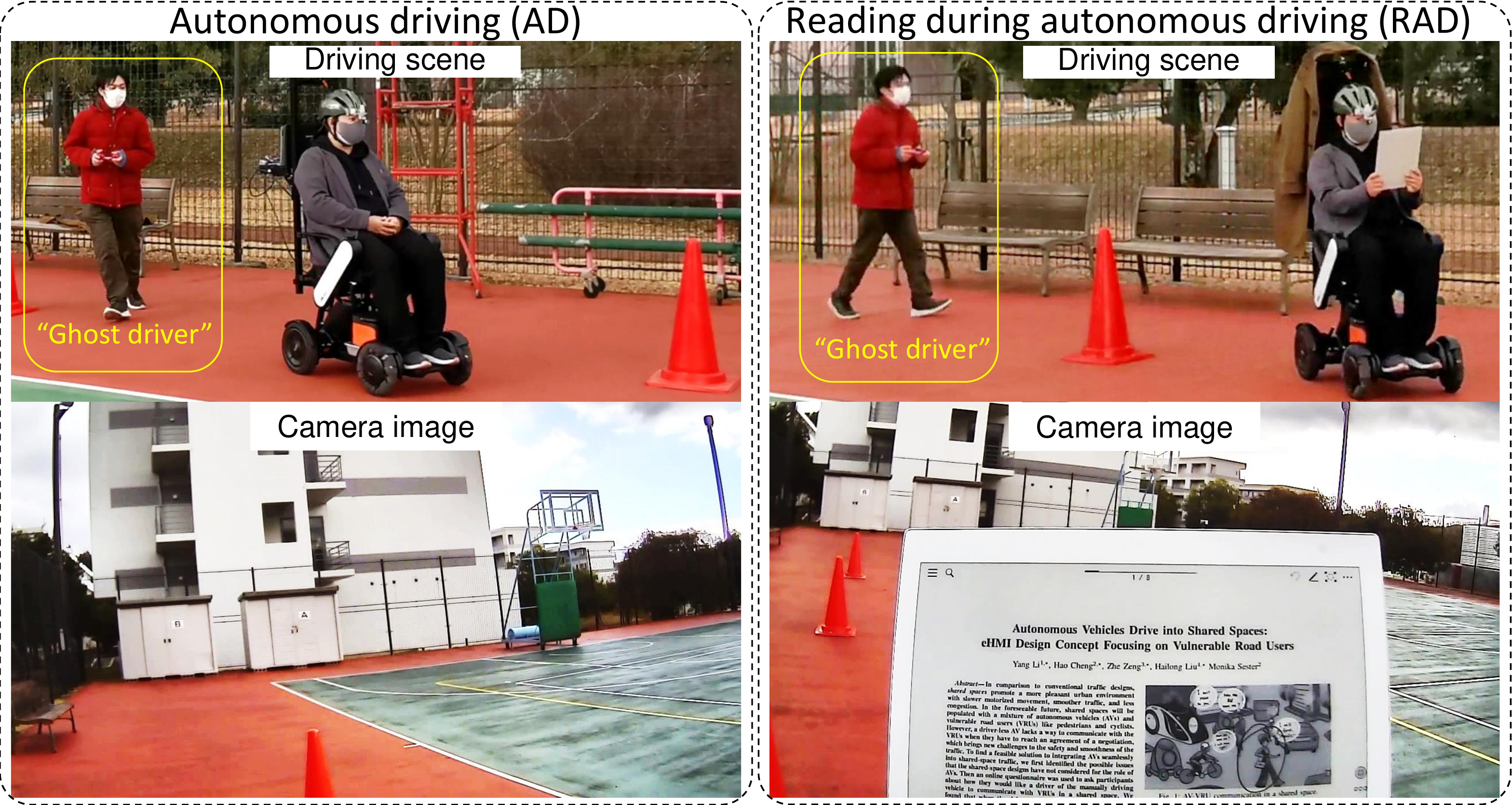}
\caption{Two driving scenarios: 1) autonomous driving (AD); 2) reading during autonomous driving (RAD).} 
\label{fig:exp02_conditions}
\vspace{-4mm}
\end{figure*}

The purpose of this experiment was to collect data to investigate how the proposed 6~DoF~SVC--VV model predicts the MSI while riding an APMV with different visual conditions: looking forward and reading-a-book, and whether it can describe the tendency that the motion sickness increases when reading books while in a moving APMV.

\subsection{Experiment design} \label{sec:exp02_desigh}
A PMV {\it WHILL Model CR} driven by remote control was used in the experiment to simulate an APMV (see the left part of Fig.~\ref{fig:exp_whill_camera}).
Its maximum speed was set to $6~km/h$, and the maximum linear acceleration was set to $1.7~m/s^2$.

We designed two experimental scenarios shown in the upper parts of Fig.\ref{fig:exp02_conditions}.
The scenarios were 1) the passenger looked ahead during autonomous driving (AD) and 2) the passenger read an e-book during autonomous driving (RAD). 
The lower part of Fig.~\ref{fig:exp02_conditions} shows the passenger's views in the two driving conditions.
During AD, the horizon and surrounding buildings were easily visible to the passenger to assist recognition of the vertical direction.
During RAD, the passenger held the e-book and moved it in sync with the head, \ie the e-book was fixed to the head coordinate system, to represent the mode of reading in which one can clearly see the text in line with the eyes.
It also was hypothesized to be more prone to causing motion sickness because the passenger may have difficulty recognizing the vertical direction because the e-book prevents the passenger from perceiving the body motion from visual information~\cite{Sato2022} such as optic flow as well as horizon or vertical.

\subsection{Driving path}

\begin{figure*}[tb] 
\centering 
\includegraphics[width=0.95\linewidth]{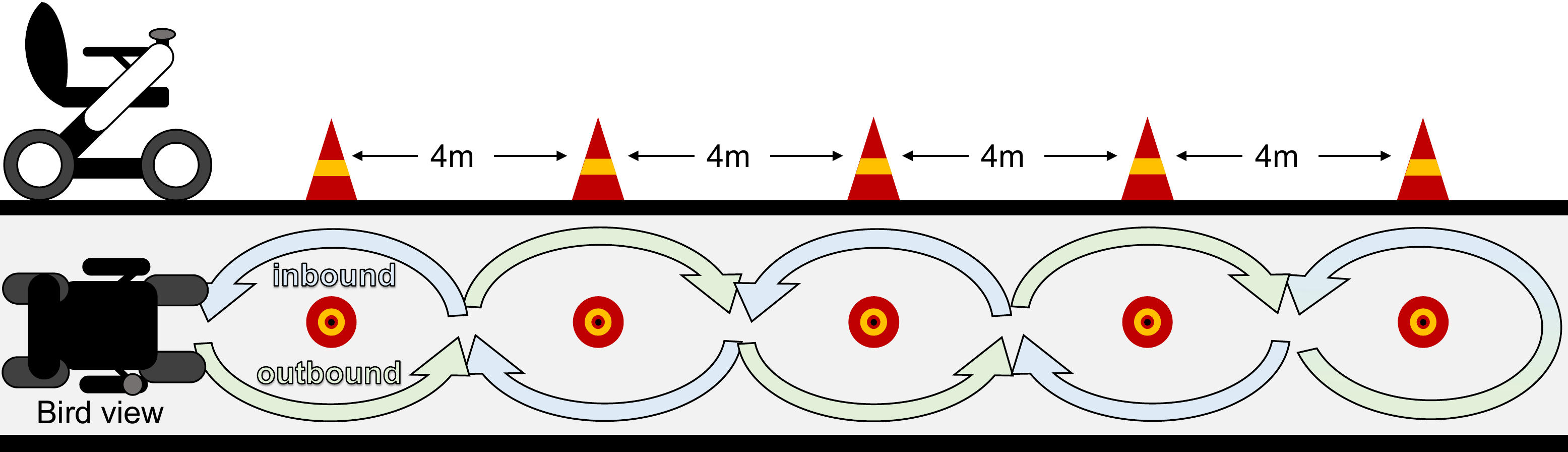}
\caption{The path designed for slalom driving.}
\label{fig:Driving_path}
\vspace{-4mm}
\end{figure*}

Leveraging the driving path design (Fig.~\ref{fig:Driving_path}) used in~\cite{wada2012can}, 
a slalom driving path is designed for this driving experiment to simulate APMV avoiding other traffic participants in mixed traffic such as shared space.

For these trials, 5 pylons were set on the road.
The pylon spacing was set to $4~m$ distance apart to account for the maximum speed of the APMV as it drove along the winding path. 
For each trial, the APMV made a U-turn past the farthest pylon to return to the starting point by slalom driving.
On this path, the passenger on the APMV may be prone to motion sickness due to frequent avoidance behaviors.

\subsection{Autonomous driving simulated by a ghost driver}

For safety, a well-trained ``ghost driver''~\cite{rothenbucher2016ghost} followed behind the PMV and drove it using a remote control to simulate APMV driving for the passenger~(Fig.\ref{fig:exp02_conditions}).

\subsection{Data measurement and preprocessing }

This experiment was conducted on a closed outdoor basketball court at NAIST, Japan. 
One passenger participated in four trials (first two: AD; second two: RAD).
The size of data measured by IMU and camera from each trial was as follows: AD~\#1: 2217 frames ($36.95~s$), AD~\#2: 2224 frames ($37.10~s$), RAD~\#1: 2333 frames ($38.88~s$), RAD~\#2 2383 frames ($39.72~s$).

The purpose of this experiment was to investigate how the proposed 6~DoF~SVC--VV model describes the phenomenon of increased motion sickness from IMU and camera data when reading a book while driving in a moving APMV.
Therefore, the passenger was not required to report any subjective response including motion sickness, which allows us to shorten the time duration of driving trials. 
Given the short--time duration of collected data, data from each trial was duplicated 10 times to calculate MSI. 

\subsection{Evaluation method}

\begin{figure*}[tb]
\centering
\begin{minipage}[t]{0.345\linewidth}
\centering
\includegraphics[width=1\linewidth]{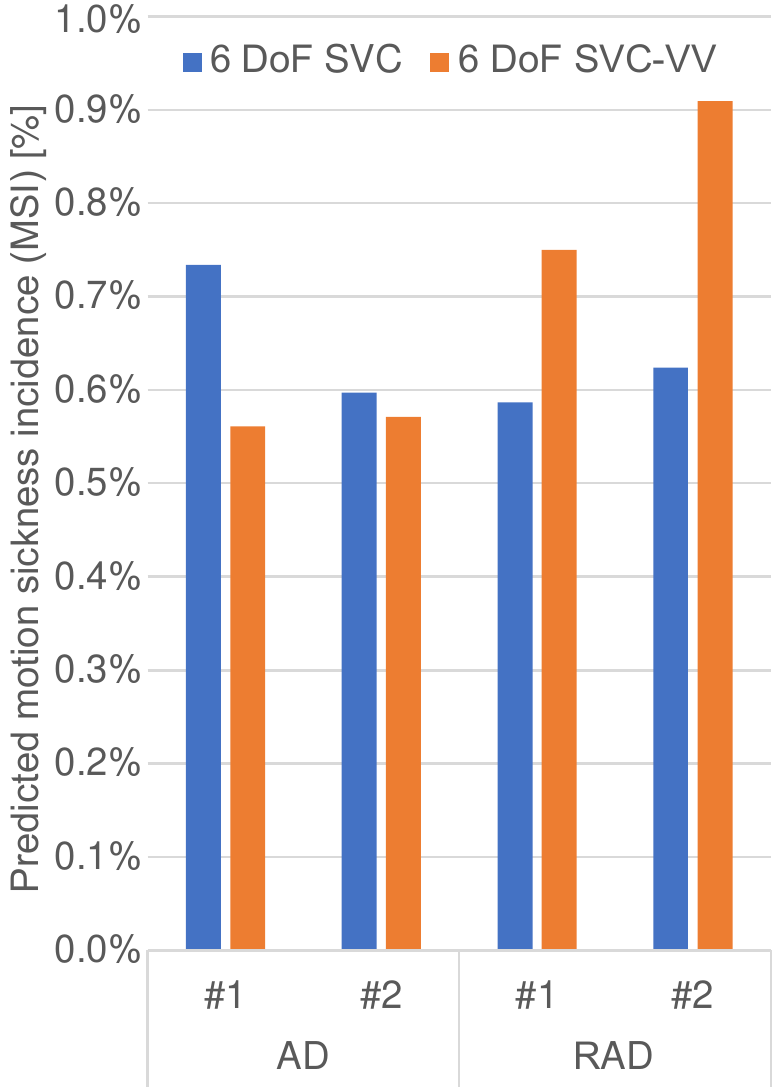}
\caption{Final MSIs after each trial repeated 10 times.}
\label{fig:MSI_final}
\end{minipage}
\hspace{1mm}
\begin{minipage}[t]{0.635\linewidth}
\centering
\includegraphics[width=1\linewidth]{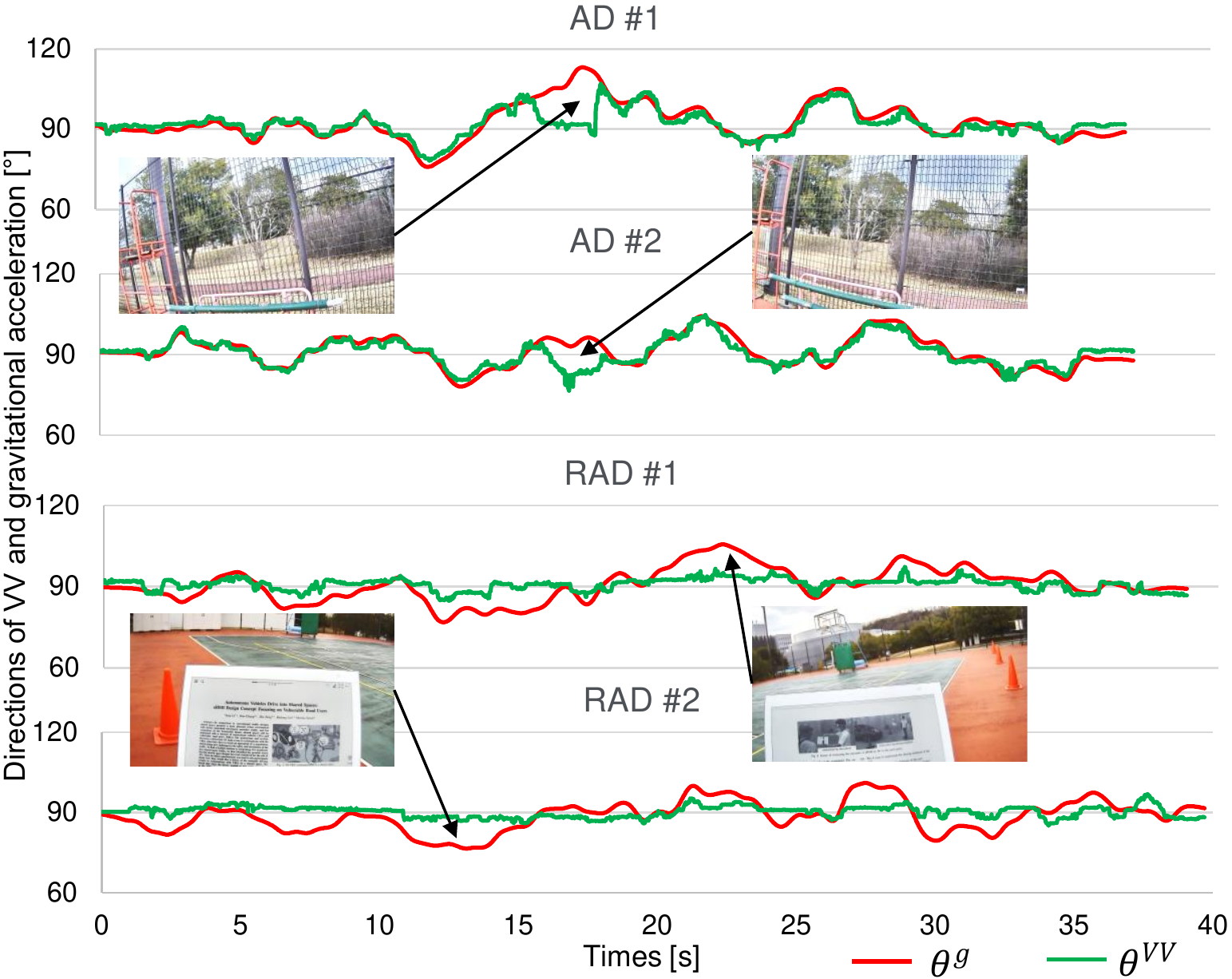}
\caption{Directions of gravitational acceleration and VV during each trial without duplicates.} 
\label{fig:exp_02_g_VV}
\end{minipage}
\vspace{-5mm}
\end{figure*}

The proposed 6~DoF~SVC--VV model was used to calculate the MSI from data measured in the four trials.
Besides, the conventional 6~DoF SVC model~\cite{wada2018analysis} was used as the baseline.
The parameters of both models were set to those found in~\cite{wada2018analysis}, apart from $K_{vvc}$ (see Table~\ref{tab:parameter}).
For the conventional 6~Dof SVC model, $K_{vvc}=0.0$, because it does not include the VV input. 
For the 6~DoF~SVC--VV model, the $K_{vvc}=5.0$ as same as $K_{vc}$ in order to balance the feedback strength of the two integrated error signal.


\begin{table}[t]
\centering
\footnotesize
\setlength\tabcolsep{1.2pt}
\caption{Model parameters}
\label{tab:parameter}
\begin{tabular}{@{}c|ccccccccccc@{}}
\toprule
Model & $K_a$ & $K_{\omega}$ & $K_{\omega c}$ & $K_{ac}$ & $K_{vc}$ & $K_{vvc}$ & \begin{tabular}[c]{@{}c@{}}$\tau$\\ $[s]$\end{tabular} & \begin{tabular}[c]{@{}c@{}}$\tau_{d}$\\ $[s]$\end{tabular} & \begin{tabular}[c]{@{}c@{}}$b$\\ $[m/s^2]$\end{tabular} & \begin{tabular}[c]{@{}c@{}}$\tau_{I}$\\ $[min]$\end{tabular} & $P$ \\ \midrule
6~DoF SVC & 0.1 & 0.8 & 10 & 1.0 & 5.0 & \textbf{0.0} & 5.0 & 7.0 & 0.5 & 12 & 85 \\ 
6~DoF~SVC--VV & 0.1 & 0.8 & 10 & 1.0 & 5.0 & \textbf{5.0} & 5.0 & 7.0 & 0.5 & 12 & 85 \\ \bottomrule
\end{tabular}
\vspace{-6mm}
\end{table}

\subsection{Results}

The calculated MSIs after each trial predicted by using the conventional 6~DoF SVC model and the 6~DoF~SVC--VV model are shown in Fig.~\ref{fig:MSI_final}.
The MSI calculated by the conventional 6~DoF SVC model at the end of AD~\#2, RAD~\#1, and RAD~\#2 were similar; however, that of AD~\#1 was higher than others. 
In contrast, using the 6~DoF~SVC--VV model, the MSIs of RAD~\#1 and RAD~\#2 were higher than MSIs of AD~\#1 and AD~\#2.
Significance test of the difference was not conducted because only one subject participated.
Those results showed that the result of 6~DoF~SVC--VV model matched the assumption of this experiment, whereas those of the conventional 6~Dof SVC model did not.

\subsection{Discussion}

To determine the causes of the above results, the directions of VV and gravitational acceleration were investigated.
Gravitational acceleration $\bm{g}$ can be calculated from the angular velocity $\bm{\omega}$ using an updated law $d\bm{g}/dt = -\bm{\omega} \times \bm{g}$~\cite{kamiji2007modeling}.
The initial $\bm{g}_0$ was set to the initial GIA, $\bm{f}_0$, measured by IMU because the initial state of the passenger's head was static ($\bm{a}_0$=0, then $\bm{f}_0=\bm{g}_0$).

As shown in Fig.~\ref{fig:exp_02_g_VV}, the directions of gravitational acceleration and VV without overlap in each trial are indicated by red and green lines.
The MAD between them were evaluated for each trial: AD~\#1: 2.57\degree, AD~\#2: 1.96\degree, RAD~\#1: 4.07\degree and RAD~\#2: 4.79\degree.
Based on those results, the directions of VV and gravitational acceleration were similar most of the time for AD~\#1 and AD~\#2.
In contrast, they were dissimilar most of the time for RAD~\#1 and RAD~\#2.

From AD~\#1 and AD~\#2, the camera images in Fig.~\ref{fig:exp_02_g_VV} represent the directions of VV and gravitational acceleration. 
First, it was difficult to identify the horizon from the camera images because the slalom track was too close to a fence when the PMV circled the farthest pylon. 
Second, the wide range of leaves in the images may impact the VV estimation.
These problems could be solved by increasing the camera's angle-of-view which should be determined based on the human motion perception characteristics.
Or, human might also have poor VV accuracy in such cases.

Referring to the camera images of RAD~\#1 and RAD~\#2, we found that the e-book obscured most of the background, causing the VV method to have difficulty detecting horizontal features from images. 
Meanwhile, the VV estimation result was strongly affected by the e-book's framework because the framework's three borders always remained horizontal or vertical in passenger's view.
It led to the directions of VV changed less in RAD~\#1 and RAD~\#2 than in AD~\#1 and AD~\#2, \ie the standard deviations~(SD) of AD~\#1 and AD~\#2 were 4.83\degree and 5.11\degree but SD of RAD~\#1 and RAD~\#2 were 2.15\degree and 2.06\degree.
When the passengers are unclear about the vertical direction (\eg the line-of-sight is blocked), they may look for vertical clues from objects with straight sides. 
The calculated results reflect this assumption.

In summary, the results obtained in the present study implies that the proposed VV–SVC model has an ability to describe the difference of the severity of motion sickness for different visual vertical conditions such as increase of motion sickness when reading books during APMV while the conventional 6~Dof SVC model~\cite{kamiji2007modeling} does not.

\section{CONCLUSION}

This study proposed a new computational model of the SVC theory for motion sickness prediction that considers the interaction between the VV from visual system and the sensed vertical signal from vestibular system.
A simple VV estimation method was proposed based on image processing, and it was added to the conventional 6~DoF SVC model~\cite{wada2018analysis} as a visually perceived vertical block.

The static experiment demonstrated that the VV estimation method accurately predicts vertical direction with low MAD.
Furthermore, the driving experiment demonstrated that the proposed 6~DoF~SVC--VV model more accurately predicts  MSI than the conventional 6~DoF SVC model when the directions of VV and gravitational acceleration differ, such as when a passenger reads a book while using an APMV.
This implies that the proposed model has a good performance to model the difference of sickness severity due to different visual vertical conditions.

\subsection{Limitations}
The VV estimation method proposed in this paper can only predict 2D VV in the head coordinate system.
Therefore, VV changes caused by head rotations on the pitch axis cannot be calculated, yet.

The camera lens' angle-of-view used in the experiment is smaller than the human angle-of-view.
This may have led to differences between the VV estimated by the proposed method, and that of a human.

The parameters in Table~\ref{tab:parameter} were referred from the conventional 6~DoF SVC model of a previous study~\cite{kamiji2007modeling}, which did not include visual--vestibular interaction.
Thus, these parameters probably may not be the optimal parameters for the proposed 6~DoF~SVC--VV model.

Only one participant participated in the experiment, as a passenger in the APMV. 
Personal habits of head behaviors that resist lateral acceleration may have affected the results.

\subsection{Future works}
Building algorithms to predict 3D VV using optical flow method with a wide-angle-lens camera is an important future work.
6~DoF~SVC--VV model parameter optimization should also be conducted.
In addition, a large-scale subject experiment with longer motion exposures to collect severity of motion sickness should be performed to confirm accuracy of the 6~DoF~SVC--VV model from the perspective of motion sickness severity.

Moreover, our previous 6~DoF SVC model including visual flow~\cite{wada2020computational} well-described the effects of angular velocity estimated from camera images.
In the RAD scenario, the e-book might have interfered with the optical flow needed to calculate angular velocity because the e-book and the head were relatively still.
We conjecture that a similar trend of motion sickness as that shown in Fig.~\ref{fig:MSI_final} may be obtained using our previous 6~DoF SVC model including visual flow~\cite{wada2020computational}.
Therefore, integration of the proposed 6~DoF~SVC--VV model and 6~DoF SVC model including visual flow is also an important future direction.

\section*{ACKNOWLEDGMENTS}
This work was supported by JST A-STEP Grant Number JPMJTR20RR and JSPS KAKENHI Grant Number 21K18308, Japan. 
We thank Prof. Luis Y. Morales for his technical support of the WHILL RC control system.

\bibliographystyle{ieeetr}
\bibliography{sample.bib}

\end{document}